\documentclass[12pt]{iopart}
\usepackage[numbers,sort&compress]{natbib}
\expandafter\let\csname equation*\endcsname\relax
\expandafter\let\csname endequation*\endcsname\relax
\usepackage{amsmath}
\usepackage{microtype} 
\usepackage{amssymb}
\usepackage{amsthm}
\usepackage[ruled,vlined,linesnumbered]{algorithm2e}
\usepackage{arydshln}
\usepackage{array}
\usepackage{amssymb}
\usepackage{dcolumn}
\usepackage{bm}
\usepackage[colorlinks, pdfborder={0 0 0}, plainpages=false]{hyperref}
\usepackage{graphicx}
\usepackage{xcolor}
\usepackage{arydshln}
\usepackage{xspace}
\usepackage{color}
\usepackage[normalem]{ulem}
\usepackage{lineno}

\usepackage{letltxmacro}
\usepackage{amsfonts}
\usepackage{amssymb}
\usepackage{bm}
\usepackage{dcolumn}
\usepackage{epsfig}
\usepackage{graphics}

\usepackage{latexsym}
\usepackage{rotating}
\usepackage{hyperref}
\usepackage{float}
\usepackage{xspace} 
\usepackage{mathrsfs}
\usepackage{subfigure}
\usepackage{multirow}
\usepackage{booktabs}
\usepackage{soul}
\usepackage{xcolor}

\usepackage{colortbl}
\usepackage{etoolbox}

\def\lesssim{\mathrel{\hbox{\rlap{\hbox{\lower4pt\hbox{$\sim$}}}\hbox{$<$}}}}
\def\gtrsim{\mathrel{\hbox{\rlap{\hbox{\lower4pt\hbox{$\sim$}}}\hbox{$>$}}}}
\def\alt{\mathrel{\hbox{\rlap{\hbox{\lower4pt\hbox{$\sim$}}}\hbox{$<$}}}}
\def\agt{\mathrel{\hbox{\rlap{\hbox{\lower4pt\hbox{$\sim$}}}\hbox{$>$}}}}

\usepackage{mathtools}

\def\gta{\ifmmode {\mathbin{\lower 3pt\hbox   
    {$\,\rlap{\raise 5pt\hbox{$\char'076$}}\mathchar"7218\,$}}}
    \else {${\mathbin{\lower 3pt\hbox
    {$\rlap{\raise 5pt\hbox{$\char'076$}}\mathchar"7218\,$}}}
    $}\fi}
\def\lta{\ifmmode {\,\mathbin{\lower 3pt\hbox   
    {$\,\rlap{\raise 5pt\hbox{$\char'074$}}\mathchar"7218\,$}}}
    \else {${\mathbin{\lower 3pt\hbox
    {$\rlap{\raise 5pt\hbox{$\char'074$}}\mathchar"7218\,$}}}
    $}\fi}

\newcommand{\msun}{{\rm M}_{\odot}}
\newcommand{\beq}{\begin{equation}}
\newcommand{\eeq}{\end{equation}}
\newcommand{\bea}{\begin{eqnarray}}
\newcommand{\eea}{\end{eqnarray}}

\definecolor{darkperiwinkle}{RGB}{102, 102, 128}

\definecolor{light-gray}{gray}{0.9}

\begin{document}

\title{Statistically-informed deep learning for gravitational wave parameter estimation}

\author{Hongyu Shen$^{1, 2}$, E. A. Huerta$^{3, 4}$,
Eamonn O'Shea$^5$, Prayush Kumar$^{5, 6}$, and Zhizhen Zhao$^{1, 2, 7}$}


\address{$^1$Department of Electrical and Computer Engineering, University of Illinois at Urbana-Champaign, Urbana, Illinois 61801, USA}
\address{$^2$Coordinated Science Laboratory, University of Illinois at Urbana-Champaign, Urbana, Illinois 61801, USA}
\address{$^3$Data Science and Learning Division, Argonne National Laboratory, Lemont, Illinois 60439, USA}
\address{$^4$Department of Computer Science, University of Chicago, Chicago, Illinois 60637, USA}
\address{$^5$Cornell Center for Astrophysics and Planetary Science, Cornell University, Ithaca, New York 14853, USA}
\address{$^6$International Centre for Theoretical Sciences, Tata Institute of Fundamental Research, Bangalore 560089, India}
\address{$^7$National Center for Supercomputing Applications, University of Illinois at Urbana-Champaign, Urbana, Illinois 61801, USA}


\begin{abstract}
\noindent We introduce deep learning models to estimate 
the masses of the binary components of black hole 
mergers, \((m_1,m_2)\),
and three astrophysical properties of the post-merger 
compact remnant, namely, the final spin, \(a_f\), and 
the frequency 
and damping time of 
the ringdown oscillations of the fundamental \(\ell=m=2\) 
bar mode, \((\omega_R, \omega_I)\). Our neural networks 
combine a 
modified \texttt{WaveNet} architecture with 
contrastive 
learning and normalizing flow. We validate 
these models against a Gaussian conjugate prior 
family whose posterior distribution is described by a 
closed analytical expression. Upon confirming that 
our models produce statistically consistent results, 
we used them to estimate the astrophysical parameters 
\((m_1,m_2, a_f, \omega_R, \omega_I)\) of five 
binary black holes: \texttt{GW150914}, \texttt{GW170104}, 
\texttt{GW170814}, \texttt{GW190521} and \texttt{GW190630}. 
We use \texttt{PyCBC Inference} to directly compare 
traditional Bayesian methodologies for parameter estimation 
with our deep learning based posterior distributions. Our 
results show that our neural network models predict 
posterior distributions that encode physical 
correlations, and that our data-driven 
median results and 90\% confidence intervals 
are similar to those produced with gravitational wave 
Bayesian analyses. This methodology requires a single 
V100 \texttt{NVIDIA} GPU 
to produce median values and posterior distributions within 
two milliseconds for each event. This neural network, and a tutorial for its use, are available at the \texttt{Data and Learning Hub for Science}.
\end{abstract}

\maketitle

\section{Introduction}

The advanced LIGO~\cite{DII:2016,LSC:2015} and advanced 
Virgo~\cite{Virgo:2015} observatories have reported 
the detection of tens of gravitational wave sources~\cite{o1o2catalog,Abbott:2020niy,Abbott:2020gyp}. 
At design sensitivity, these instruments will be 
able to probe a larger volume of space, thereby increasing 
the detection rate of sources populating the gravitational wave 
spectrum. Thus, given the expected scale of gravitational wave 
discovery in upcoming observing runs, it is in order to 
explore the use of computationally efficient signal-processing 
algorithms for gravitational wave detection and parameter estimation.

The rationale to develop scalable and computationally efficient 
signal-processing tools is apparent. Advanced gravitational wave 
detectors will be just one of many large-scale science 
programs that will be competing for access to oversubscribed and finite  
computational resources~\cite{HuertaBWOSG,HuertaES,2017Weitzel,nn_k_2010}. Furthermore, transformational breakthroughs 
in multi-messenger astrophysics over the next decade will be enabled by 
combining observations in the gravitational, electromagnetic 
and astro-particle spectra. The combination of these high dimensional, 
large volume and high speed datasets in a timely and innovative manner 
presents unique challenges and opportunities~\cite{whitepaper:SCIMMA,eliuMMA:2019,Huerta:2019rtg}. 

The realization that companies such as Google, YouTube, 
among others, 
have addressed some of the big-data challenges we are facing in 
multi-messenger astrophysics has motivated a number of researchers 
to learn what these companies have done, and how such 
innovation may be adapted in order to maximize the science reach of 
big-data projects. The most successful approach to date consists 
of combining deep learning with innovative and extreme scale 
computing. 

Deep learning was first proposed as a novel signal-processing tool 
for gravitational wave astrophysics in~\cite{geodf:2017a}. That initial 
approach considered a 2-D signal manifold for binary black hole 
mergers, namely the masses of the binary components \((m_1,\,m_2)\), 
and considered simulated advanced LIGO noise. The fact that 
such method was as sensitive as template-matching algorithms, but 
at a fraction of the computational cost and orders of magnitude faster, 
provided sufficient motivation to extend such methodology and apply 
it to detect real gravitational wave sources in advanced LIGO 
noise in~\cite{geodf:2017c,GEORGE201864}. These studies have sparked 
the interest of the gravitational wave community to explore 
the use of deep learning for the detection of the large zoo of gravitational wave sources~\cite{2018GN,Skliris:2020qax,Lin:2020aps,Wang:2019zaj,Nakano:2018vay,Fan:2018vgw,Li:2017chi,Deighan:2020gtp,Miller:2019jtp,Krastev:2019koe,2020PhRvD.102f3015S,Dreissigacker:2020xfr,Khan:2020foe,Dreissigacker:2019edy,Wei_Hubble,2020PhRvD.101f4009B,2020arXiv200914611S,Khan:2020fso,PhysRevLett.122.211101,Rebei:2018R,wei_warning,wei_ecc_princ}.

Deep learning methods have matured to now cover a 
4D signal manifold that describes the masses of the binary 
components and the \(z\)-component of the 3-D spin 
vector: \((m_1, m_2, s_1^z, s_2^z)\)~\cite{Open_GW_DLHub,DL_ensembles_GWs}. These algorithms 
have been used to search for and find gravitational wave sources 
processing open source advanced 
LIGO data in bulk, which is available at the \texttt{Gravitational Wave Open Science Center}~\cite{Vallisneri:2014vxa}. In the context of 
multi-messenger sources, deep learning has been used to forecast 
the merger of binary neutron stars and black hole-neutron star systems~\cite{wei_warning,2021arXiv210409438Y}. 
The importance of including 
eccentricity for deep learning forecasting has also been studied and 
quantified~\cite{wei_ecc_princ}. In brief, deep learning research is 
moving at an incredible pace. 

Another application area that has gained traction is the use 
of deep learning for gravitational wave parameter estimation.
The established approach to estimate the astrophysical parameters 
of gravitational wave signals is through 
Bayesian 
inference~\cite{bambi:2012MNRAS,bambiann:2015PhRvD,Singer_Price_2016,2019PASP131b4503B}, 
which is a well tested and extensively used method, though computationally-intensive. 
On the other hand, given the scalability and computational 
efficiency of deep learning models, the gravitational wave parameter estimation can take advantage of its power to produce faster inference.

Gravitational wave parameter estimation has 
rapidly evolved from point-wise parameter estimation~\cite{geodf:2017a,geodf:2017c,GEORGE201864} 
to the use of neural networks dropouts to provide estimation intervals~\cite{alvares2021exploring}, and to output a 
parametrized approximation of the corresponding posterior 
distribution~\cite{Chua:2019wwt}. Other methods have proposed 
the use of 
Conditional Variational Auto-Encoders (CVAEs) to infer 
the parameters of GWs embedded in simulated noise~\cite{Gabbard:2019rde,Green:2020hst}. In~\cite{green2020complete} the authors harnesses 
new methods, e.g., normalizing flow~\cite{grover2018flow}, 
to do parameter 
estimation over the full 
15-dimensional space of binary black hole system parameters for 
the event GW150914. Building upon this study, 
authors in~\cite{2021arXiv210612594D} presented deep learning 
methods to estimate the astrophysical parameters of several gravitational wave events. One can also refer to~\cite{cuoco2020enhancing,Huerta:2021ybd} for a comprehensive review of the gravitational-wave-based machine learning approaches.

In this article we quantify the ability of deep learning 
to estimate the masses of the binary components 
of binary black hole mergers, and of the astrophysical 
parameters that describe the properties of the black hole 
remnant, namely, the final spin, \(a_f\), and the frequency and 
damping time of the ringdown oscillations of the fundamental 
\(\ell=m=2\) bar mode, \((\omega_R,\,\omega_I)\), known as quasinormal modes (QNMs)~\cite{Berti:2006b}. An existing approach proposes to use neural networks to solve differential equations for QNMs~\cite{ovgun2021quasinormal}. Our approach, on the other hand, 
differs from this or other studies in the literature in that we 
estimate the astrophysical parameters 
of the remnant by directly feeding time-series 
advanced LIGO strain data into our deep learning 
algorithms.

This article is organized as follows. In Section~\ref{method} 
we describe the architecture of our neural network model, 
and the datasets used to train, validate and test it. We 
briefly describe the Bayesian inference pipeline, 
\texttt{PyCBC Inference}, in Section~\ref{lal_inf}, 
which we used as a baseline to compare the full posterior 
distributions predicted by our deep learning model. We quantify 
the accuracy and physical consistency of the predictions of our 
deep learning model for several gravitational wave sources 
in Section~\ref{experiments}. We summarize our findings and 
future directions of work in Section~\ref{conclusion}.

\section{Methods}
\label{method}

Herein we describe several methods to improve the training performance and model accuracy of our algorithms. We have used \texttt{PyTorch}~\cite{paszke2017automatic} to design, train, validate and test our neural network models. 

\subsection{Deep Learning Model Objective}
\label{related_work}

The goal of our deep learning model is to estimate a 
posterior distribution of the physical parameters of the 
waveforms from the input noisy data.  This approach 
shares similarities with Bayesian approaches such as Markov Chain Monte Carlo (MCMC), e.g., 
once a likelihood function and a predefined prior are 
provided, posterior samples may be drawn. The difference 
between the deep learning model and MCMC is that our 
proposed framework will learn a distribution from which 
we can easily draw samples, thereby increasing 
computational efficiency significantly. It is worth 
emphasizing that once the likelihood model is properly 
defined, the framework we introduce here may be applicable 
to other disciplines.

In the context of gravitational waves, the noisy waveform 
$y$ is generated according to the following physical model,

\begin{equation}
y_{i, \ell} = F(x_i) + n_{i, \ell},
\label{eq:signal_model}
\end{equation}

\noindent where $F$ is the function that maps the physical 
parameters (masses and spins) $x_i$ to the 
gravitational waveform template~\cite{2019PASP131b4503B, alexnitz20203904502, Vousden_2015}, and $n_{i, \ell}$ denotes the additive noise 
at various signal-to-noise ratios (SNR). We use $y_{i, \ell}$ 
with subscript pair $(i, \ell)$ to specifically indicate 
the $i$-th template associated with $\ell$-th noise realization 
in our dataset $D$. For simplicity, we use $y$ and $x$ to 
indicate noisy waveforms and the physical parameters when 
the specification of $i$ or $\ell$ subscript is not needed. 
We use $K$ and $M$ to denote the dimension of $y$ and $x$, respectively. 

We use \texttt{WaveNet}~\cite{oord2016wavenet} to extract 
features from the input noisy waveforms.  \texttt{WaveNet} 
was first introduced as an audio synthesis tool to 
generate human-like audios given random inputs. It uses 
dilated convolutional kernel and residual network to capture 
the spatial information both in the time domain and the model 
depth, which has been shown to be a powerful tool in 
model time-series data. Previously, 
\cite{wei2020gravitational,DL_ensembles_GWs} tailored 
this architecture for gravitational wave denoising and detection. 
The encoded feature vector $h \in \mathbb{R}^L$ comes from 
an embedding function parameterized by the \texttt{WaveNet} 
weights $\omega$, $f_\omega : y \mapsto h$. In other 
words, $h = f_\omega(y)$.

Normalizing flow is a technique to transform distributions 
with invertible parameterized functions. Specifically, 
we use a conditional version of normalizing flow: conditional autoregressive 
spline~\cite{durkan2019neural,dolatabadi2020invertible,bingham2019pyro,bingham2018pyro,phan2019composable} to learn the posterior distribution on top of the 
encoded latent space by \texttt{WaveNet} encoding. and we implement it through a PyTorch-based probabilistic programming package:  \texttt{Pyro}~\cite{bingham2019pyro}.  Mathematically, we denote the invertible function  $g_{(h, \theta)}: z \mapsto x$ is parameterized by the learnable model weights $\theta$ and the encoded feature $h$. In this way, we encode dependencies of the posterior distribution on the input $y$. The random vector $z \in \mathbb{R}^M$ is drawn according to a pre-defined base distribution $p(z)$, and has the same dimension as $x$. The function $g_{(h, \theta)}(z)$ is then used to convert the base distribution $p(z)$ to the approximated posterior distribution $\hat{p}_{\omega, \theta}(x |y)$ of the physical parameters, 
\begin{align} \label{NF_eq}
    \hat{p}_{\omega, \theta} (x\vert y) &= p(z) \left\vert \det\left(\frac{\partial g_{(h, \theta)}(z)}{\partial z}\right)\right\vert^{-1},  
\end{align}
with $h = f_\omega(y)$. 

\begin{figure}
	\begin{center}
	    {
		\includegraphics[width=0.7\linewidth]{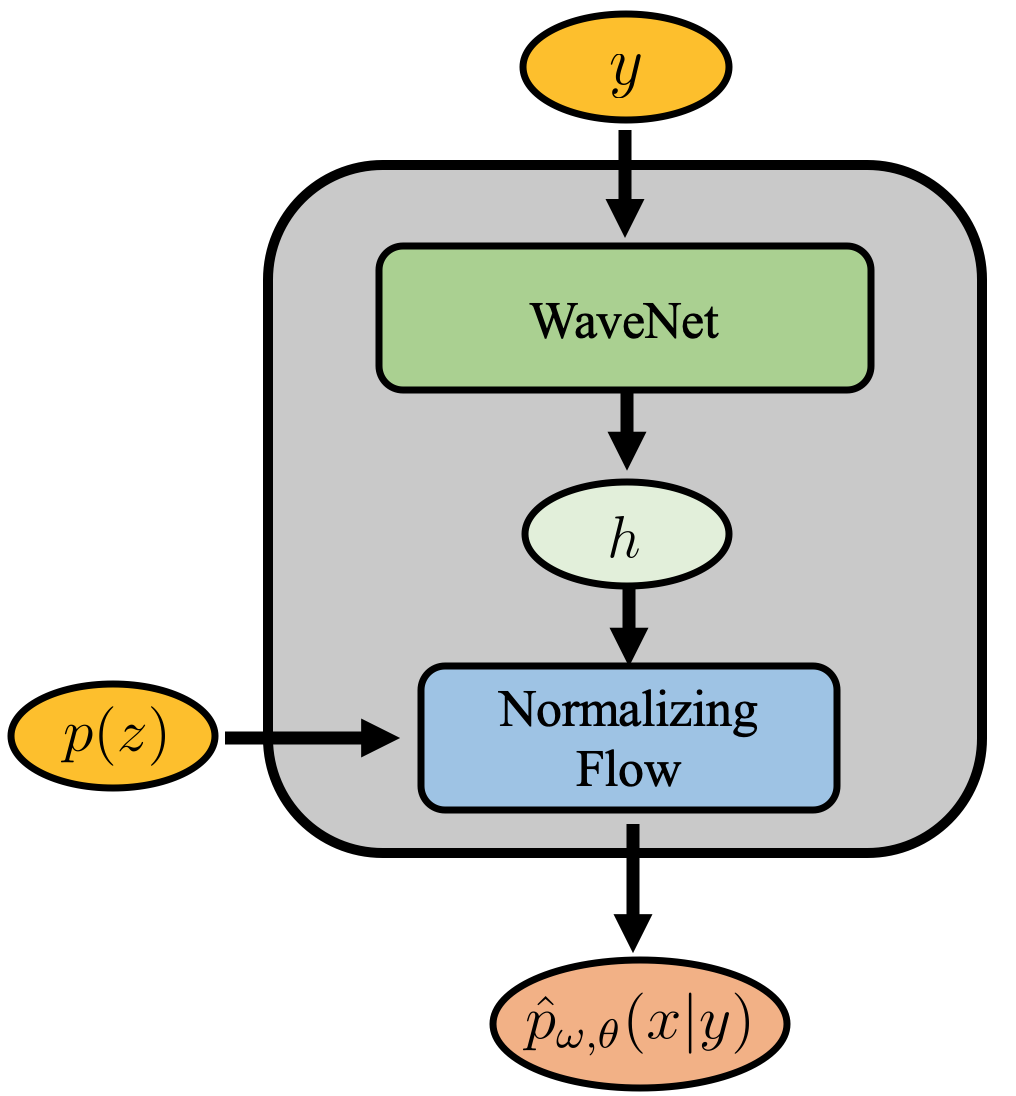}
		}
	\end{center}
	\caption{Model Architecture. The first component of our model is a \texttt{WaveNet} architecture with 11 blocks, whose input is a 1s-long waveform sampled at 4096Hz, denoted by $y$. The output of the \texttt{WaveNet} modulo is a 254 dimensional vector, $h$, that is fed into a normalizing flow modulo, which is then combined with a base distribution, $p(z)$, to provide the posterior distribution estimation $\hat{p}_{\omega, \theta} (x\vert y)$. $z$ represents the random variable for the base distribution, and $x$ represents the physical parameters of the binary black hole mergers, respectively.}
	\label{model_diagram}
\end{figure}

The computation of the transformation $g_{(h, \theta)}(z)$ contains two steps. The first step is to compute the intermediate coefficients $\alpha$ from the feature vector $h$ based on the function $k_\theta$, which is parameterized by 2 fully connected layers with weights denoted as $\theta$, i.e., $\alpha = k_\theta(h)$. The coefficients $\alpha$ are used to combine the invertible linear rational splines to form $g_{(h, \theta)}$ (see Eq. (5) in~\cite{dolatabadi2020invertible} for details). Therefore, $g_{(h, \theta)}$ is an element-wise invertible linear rational spline with coefficients $\alpha$. Since $h$ depends on the input waveform $y$ and  $\alpha = k_\theta(h)$, the resulting mapping $g_{(h, \theta)}$ and parameterized distribution in Eq.~\eqref{NF_eq} vary with the input $y$. The parameterization of the estimated posterior distribution is illustrated in Figure~\ref{model_diagram}.

To learn the network weights, we need to construct the empirical loss objective given the collection of training data $\{x_i, y_{i, \ell} \}$. We propose to include a loss term defined on the feature vectors in our learning objective to take account for the variation in the waveform due to noise. That is if the underlying physical parameters are similar, then the similarity of the feature vectors should be large,  and vice versa. To achieve this, we use contrastive learning objective~\citep{hadsell2006dimensionality} to distinguish positive data pairs (waveforms with the same physical parameters) from the negative pairs (noisy waveforms with different physical parameters). 
Specifically, we use the normalized temperature-scaled cross entropy (NT-Xent) loss used in the state-of-the-art contrastive learning technique \texttt{SimCLR}~\citep{Chen2020ASF,chen2020big}. \texttt{SimCLR} was originally introduced to  improve the performance of image classification with additional data augmentation and NT-Xent loss evaluation. We adapt the NT-Xent loss used in contrastive learning to our feature vectors, 
\begin{align}
\label{contrastive_loss2}
 l(h_{i, j}, h_{i, \ell}) \equiv - \log 
 \frac{
 e^{\text{sim}(h_{i, j}, h_{i, \ell})/\tau}}{\sum_{i' \neq i} \sum_{j = 1}^2 \sum_{\ell = 1}^2 e^{\text{sim}(h_{i, j}, h_{i', \ell})/\tau}},
 \end{align}
where $h_{i, \cdot}=f_\omega(y_{i, \cdot})$, $\tau \in (0, \infty)$ is a scalar temperature parameter, and we choose 
$\tau=0.2$ according to the default setting provided in~\cite{Chen2020ASF}. The NT-Xent loss performs in such a way that, regardless of the noise statistics, the cosine distances of the encoded features associated with the same underlying physical parameters (i.e. $h_{i,j}$ and $h_{i, \ell}$) are minimized, and the distances of  features with different underlying physical parameters are maximized. Consequently, the trained model is robust to the change of noise realizations and noise statistics.  Therefore, incorporating the term in Eq.~\eqref{contrastive_loss2} can be used as a noise stabilizer for gravitational wave parameter estimation. We found that the inclusion of this term speeds up the convergence in training.

Our deep learning objective in Eq.~\eqref{model_obj_batch} combines the NT-Xent loss in Eq.~\eqref{contrastive_loss2} with the 
posterior approximation term. Given a batch of $B$ physical parameters $x_i$, we generate different noise realizations $y_{i, \ell}$ for each $x_i$ and the empirical loss function is,   
\begin{align}
\label{model_obj_batch}
L(\omega, \theta) =    \frac{1}{2B} & \sum_{i = 1}^B \left [  -\sum_{\ell = 1}^2 \log \hat{p}_{\omega, \theta}(x_i\vert y_{i, \ell})  \right. \nonumber \\
& \left. +   \sum_{\ell = 1, \, \ell \neq j }^2\sum_{j = 1}^2 l (f_\omega(y_{i, j}), f_\omega(y_{i, \ell}))  \right],
 \end{align}
where $\hat{p}_{\omega, \theta}(x_i\vert y_{i, \ell})$ is defined in Eq.~\eqref{NF_eq}. Minimizing the loss in Eq.~\eqref{model_obj_batch} with respect to 
$\omega$ and $\theta$ provides a posterior 
estimation for gravitational wave 
events.  

It is worth pointing out that while Refs.~\cite{Dumoulin2017AdversariallyLI,Donahue2017AdversarialFL} 
apply $q(z)$, an arbitrary random distribution to their 
generative model, our posterior distributions do not involve 
arbitrary random distributions.

\subsection{Separate Models for Parameters}
\label{sec:param_sep}

In this paper, we are interested in the following physical parameters: \((m_1, m_2, a_f, \omega_R, \omega_I)\). 
We find that trying to estimate all parameters using a single model lead to sub-optimal results given that they are of different scales. Thus, we use two separate models with similar model architecture as shown in Figure~\ref{model_diagram}. One model is used to estimate the masses \((m_1, m_2)\) of the binary 
components, while the other one is used to infer the final 
spin \((a_f)\) and QNMs \((\omega_R, \omega_I)\) of the remnant. 

The final spin of the remnant and its QNMs have similar 
range of values when the QNMs are cast in dimensionless units. 
We trained the second model using the fact that the QNMs are determined by the final spin \(a_f\) using the relation~\cite{Berti:2006b}:
\begin{equation}
\omega_{220}\left(a_f\right)= \omega_R + i\, \omega_{I} ,
\label{qnms}
\end{equation} 
\noindent where \((\omega_R,\,\omega_{I})\) correspond to the frequency and damping time of the ringdown oscillations for the fundamental \(\ell=m=2\) bar mode, and the 
first overtone \(n=0\). We compute the QNMs 
following~\cite{Berti:2006b}. One can translate 
\(\omega_R\) into the ringdown frequency (in units of Hertz) and \(\omega_I\) into the corresponding (inverse) damping time 
(in units of seconds) by computing $M_f \cdot \omega_{220}$, where \(M_f\) is the final mass of the remnant, and can 
be determined using Eq.~(1) in~\cite{HealyLous:2017PRDH}. An additional benefit of using two separate models is that 
the training converges faster with two models considering two different sets of physical parameters at different magnitudes.

\subsection{Dataset Preparation and Training}
\label{method_data_prep}

\textbf{Modeled Waveforms} We used the surrogate waveform 
family~\cite{blackman:2015} to produce modeled waveforms 
that describe binary black holes with component 
masses \(m_{1}\in[10\msun,\,80\msun]\), \(m_{2}\in[10\msun,\,50\msun]\), and 
spin components \(s^z_{\{1,\,2\}}\in[-0.9,\,0.9]\).
By uniformly sampling this parameter space we produce a 
dataset with 1,061,023 waveforms. These waveforms describe 
the last second of the late inspiral, merger, and ringdown. 
The waveforms are produced using a sample rate of 4096Hz.

For training purposes, we label the waveforms using the 
masses and spins of the binary components, and then use this 
information to also enable the neural net to estimate the 
final spin of the black hole remnant using the formulae 
provided in~\cite{Hofmann:2016yih}, and the QNMs following~\cite{Berti:2006b}. 
In essence, we are training our neural network models to 
identify the key features that determine the properties of the 
binary black holes before and after merger using a unified framework.

We use 90\% of these waveform samples for training, 10\% testing. 
The training samples are randomly and uniformly chosen. 
Throughout the training, we use AdamW optimizer to minimize 
the mean squared error of the predicted parameters with 
default hyper-parameter setups~\citep{loshchilov2017decoupled}. 
We choose the learning rate to be 0.0001. To 
simulate the environment where the true gravitational waves 
are embedded, we use real advanced LIGO noise to compute power 
spectral density (PSD), which is then used to whiten the 
templates. 

\textbf{Advanced LIGO noise.} For training we used a 4096s-long 
advanced LIGO noise data segment, sampled at 4096Hz, starting at GPS 
time 1126259462. We obtained these data from 
the \texttt{Gravitational Wave Open Science Center}~\cite{Vallisneri:2014vxa}. \textcolor{black}{We estimate a PSD using the entire 4096s segment to whiten the modeled waveforms and noise. For each one second long noisy waveform used in training, we combine the clean whitened template with a randomly picked one second long noise segment from the 4096s-long advanced LIGO strain data. For each generated waveform template (see Eq.~\ref{eq:signal_model}), we apply two different noisy realizations. As a result, the total number of noisy waveforms (clean templates $+$ noise realizations) applied during training is equal to: $\#$ of training iterations $\times$ batch size $\times$ 2. }

In Section~\ref{experiments}, we demonstrate that our model, trained 
only with advanced LIGO noise from the first observing run, 
is able to estimate the astrophysical parameters of other 
events across O1-O3. We fixed the merger point of the training 
templates at the 3,596$^\text{th}$ timestep out of 4,096 total timesteps. 
We empirically found having a fixed merger point, rather 
than shifting the templates to have time-invariant property, 
provides a tighter estimation of the posteriors. Our deep 
learning model was trained on 1 \texttt{NVIDIA} V100 GPU 
with a batch size of 8. In general, it takes about 1-2 days 
to fully train this model. 

\subsection{GPS Trigger Time}
\label{sec:cro_cor}
It is known that a trigger GPS time associated with a  gravitational wave event, typically provided by a detection algorithm, may differ from the true time of coalescence.  Therefore, we perform a local search around the trigger time by any given detection algorithm as a pre-processing step for the parameter estimation using the trained model. We first identify local merger time candidates by evaluating the normalized cross-correlation (NCC) of the whitened observation with 33,713 whitened clean templates, whose physical parameters uniformly cover the range: \(m_1\in[10\msun,\,80\msun]\), \(m_2\in[10\msun,\,50\msun]\), and \(s^z_{\{1,\,2\}}\in[-0.9,\,0.9]\), over a time window of 0.015 seconds around the time candidates. The time points with top NCC values are selected as the candidates. Then we use the trained models to estimate the posterior distributions of the physical parameters at each candidate time point. \textcolor{black}{In practice, we found that the trigger times with the best NCC values differ from those published at the \texttt{Gravitational Wave Open Science Center} by up to 0.01s. These trigger times produce different posterior distributions that vary in size by up to $\pm1\msun$ for the masses of the binary components, and up to 5\% for the astrophysical properties of the compact remnant.} We have selected the time point that gives the smallest $90\%$ confidence area for the results we present in Section~\ref{real_event_results_check}.

\section{Bayesian Inference}
\label{lal_inf}

We compare our data-driven posterior estimation with \texttt{PyCBC Inference}~\cite{2019PASP131b4503B, alexnitz20203904502, Vousden_2015}, which uses a parallel-tempered MCMC algorithm, \texttt{emcee\_pt}~\cite{2013PASP..125..306F}, to evaluate the posterior probability $p(x|y)$ for the set of source parameters $x$ given the data $y$. The posterior is calculated as $p(x|y) \propto p(y|x) p(x)$ where $p(y|x)$ is the likelihood and $p(x)$ is the prior. The likelihood function for a set of $N$ detectors is 
\begin{equation} \label{likelihood} 
p(y | x) = \exp \left( -\frac{1}{2} \sum_{i=1}^N \langle \hat{y}_i(k) -  \hat{s}_i(k, x) | \hat{y}_i(k) -  \hat{s}_i(k, x) \rangle \right),
\end{equation} 
where $\hat{y}_i(k)$ and $\hat{s}_i(k, x)$ are the frequency-domain representations of the data and the model waveform for detector $i$. The inner product $\langle \cdot | \cdot \rangle$ is defined as
\begin{equation} \label{inner_product}
    \langle \hat{a}_i(k) | \hat{b}_i(k) \rangle = 4 \mathcal{R} \int_0^{\infty} 
    \frac{\hat{a}_i(k) \hat{b}_i(k) }{P^i (k)}
    \mathrm{d} k\,,
\end{equation}
\noindent where $P^i(k)$ is the PSD of the $i$-th detector. 

We performed the MCMC analysis using the publicly available data from the GWTC-1 release~\cite{o1o2catalog} and used the corresponding publicly available PSD files for each event~\cite{PSDs}. We analyse a segment of 8 seconds around the GPS trigger 1167559935.6,  with the data sampled to 2048 Hz. We use the IMRPhenomD~\cite{PhysRevD.93.044007} waveform model to generate waveform templates to evaluate the likelihood. We assume uniform priors for the component masses with $m_{\{1,2\}}\in [10\msun, 80\msun)$ and uniform priors on the component spins with $ a_{\{1,2\}} \in (-0.99, 0.99)$. We also set uniform priors on the luminosity distance with $ D_L \in [10, 4000) \mathrm{Mpc} $ and the deviation of the arrival time from the trigger time $-0.1 < \Delta t < 0.1$. We set uniform priors for the coalescence phase and the polarization angle $\phi_c, \psi \in [0, 2 \pi)$. The prior on the inclination angle between the binary's orbital angular momentum and the line of sight, $\iota$, is set to be uniform in the sine of the angle, and the right ascension and declination have priors to be uniform over the sky.

Furthermore, they may be used to 
cross validate the physical reality of an event~\cite{Open_GW_DLHub,DL_ensembles_GWs}, 
and to assess whether the estimated merger time is consistent 
between the two separate models. For instance, if the models 
output very different merger times, then we may 
conclude that they are not providing a reliable merger time. 
On the other hand, when their results are consistent, within 
a window between 0.001s and up to 0.005s, then 
we can remove the ambiguity introduced when using the 
NCC approach described in Section~\ref{sec:cro_cor}.

\section{Experimental Results}
\label{experiments}

In this section we present two types of results. First, we validate 
our model with a well known statistical model. Upon 
confirming that our deep learning approach is statistically 
consistent, we used to estimate the parameters of five binary black 
hole mergers.

\subsection{Validation on Simulated Data }

\begin{figure*}
	\centering
	\subfigure[]{
		\includegraphics[width=0.30\linewidth]{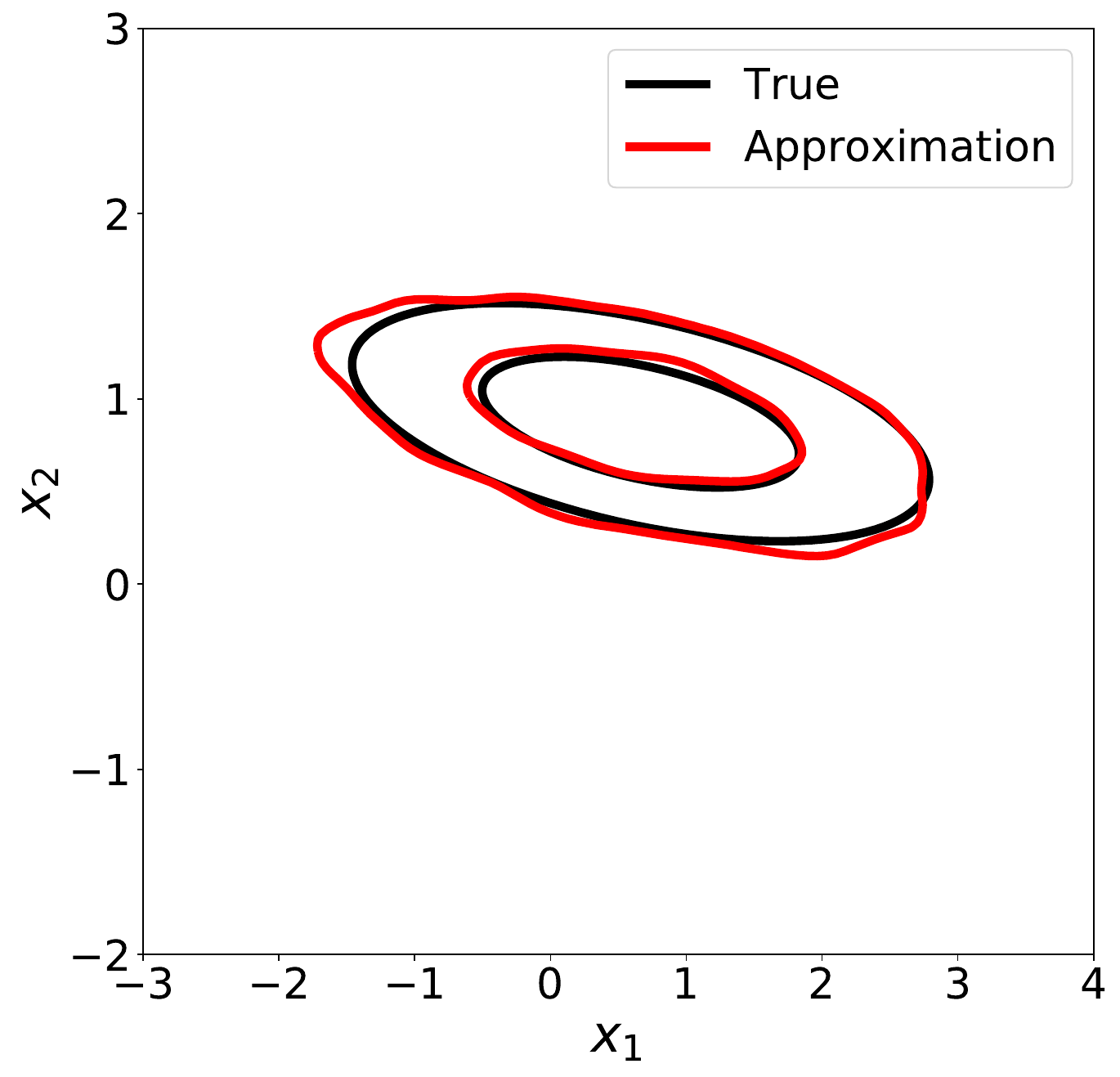}}
	\subfigure[]{
		\includegraphics[width=0.30\linewidth]{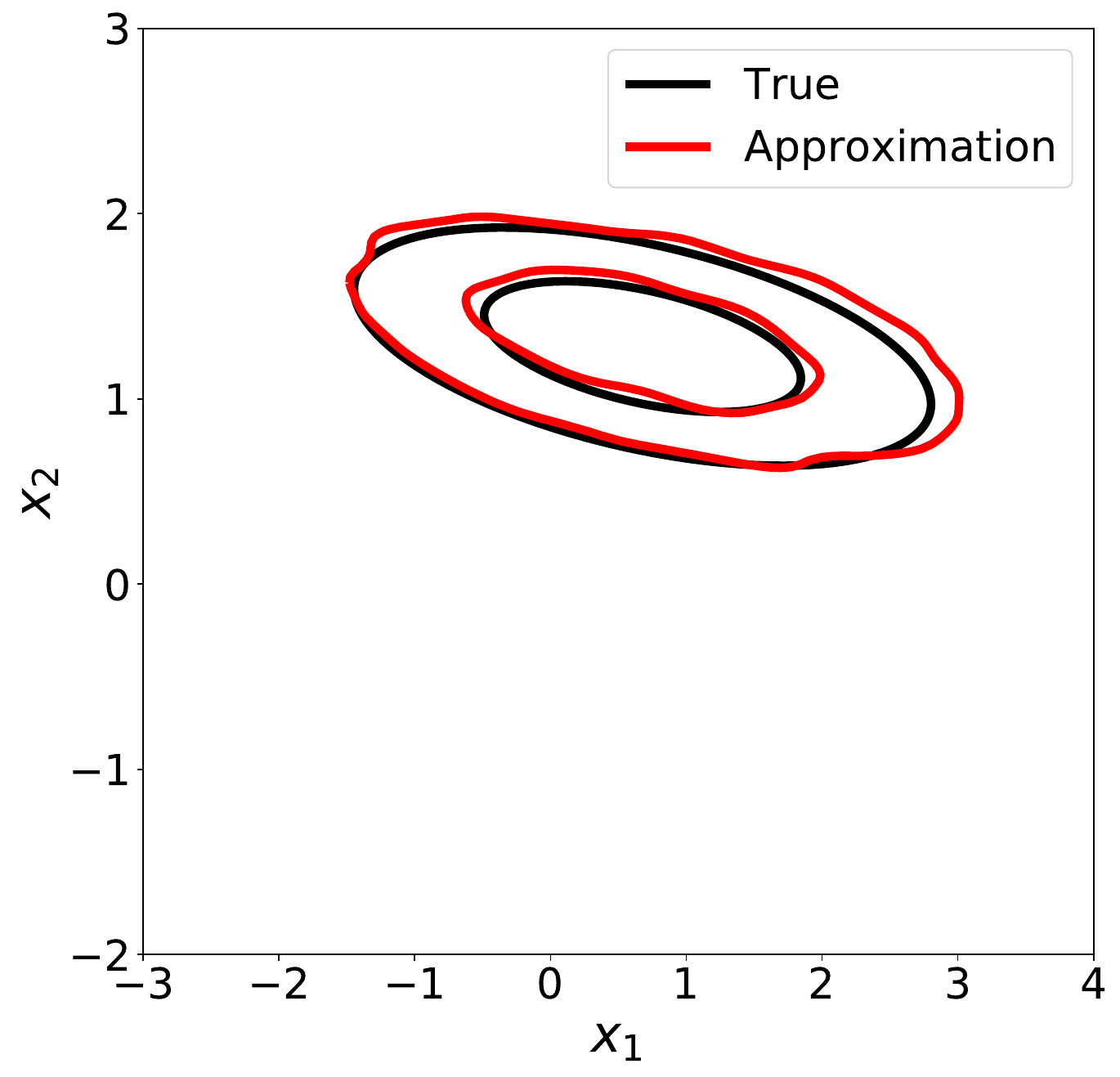}}
		\subfigure[]{
		\includegraphics[width=0.30\linewidth]{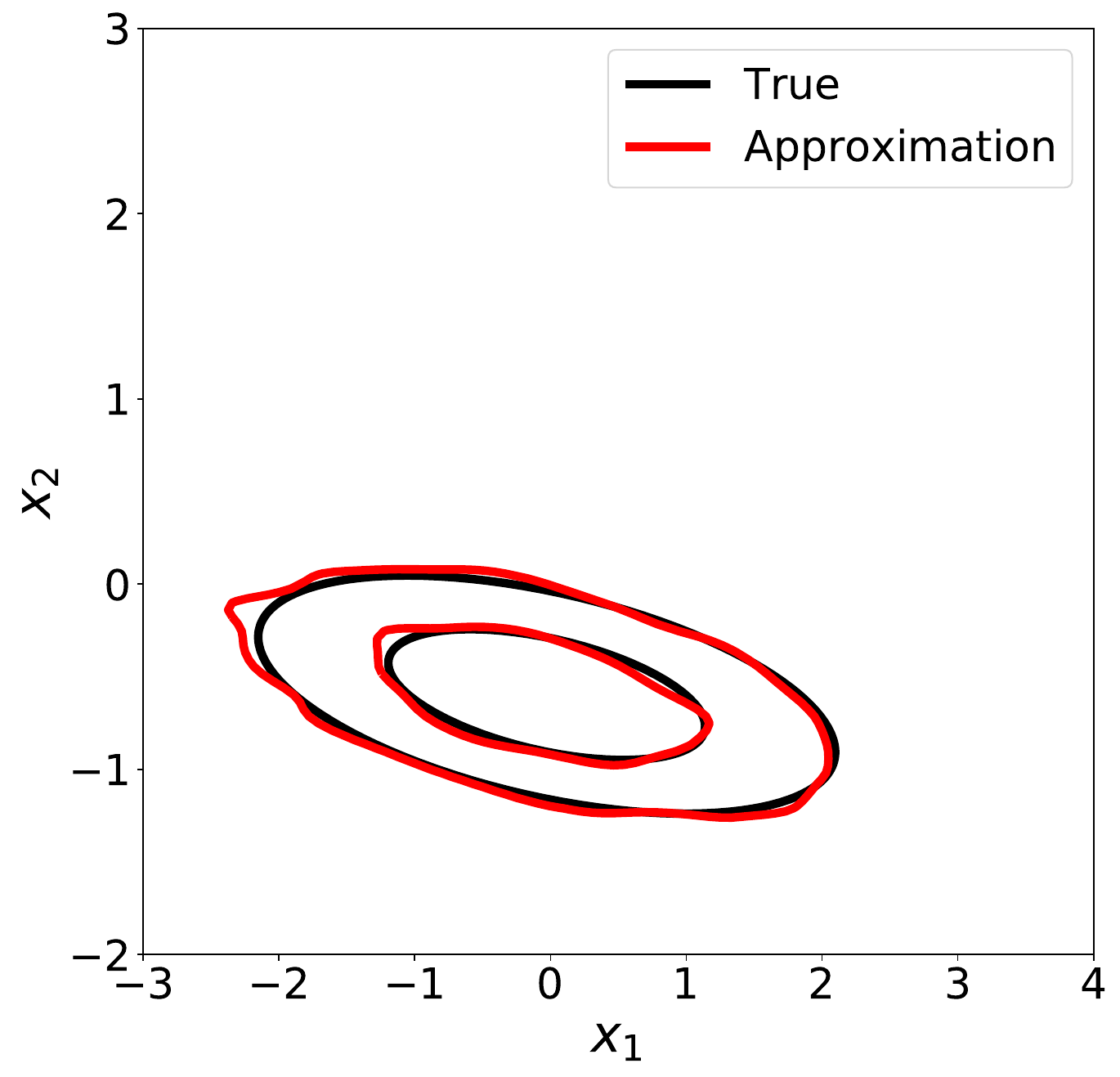}}
	\caption{Comparison of posterior distributions produced by our deep learning model (red contours) and a Gaussian conjugate prior family whose posterior distribution (black contours) is given by a closed analytical model. These data-driven predictions for the 50\% (the inner ellipse )and 90\% (the outer ellipse) confidence contours are in agreement with expected statistical results.}
	\label{sim_no_variable_fig}
\end{figure*}

We performed experiments on simulated data that have closed 
form posterior distributions. This is important to ascertain the 
accuracy and reliability of our method. The simulated data are generated through a linear observation model with additive white Gaussian noise,
\begin{equation}
\label{eq:linear_obs}
    y = Ax  + n,
\end{equation}
\noindent where  the additive noise $n \sim \mathcal{N}( \bm{0}, \sigma^2 I)$. We consider the underlying parameters $x \in \mathbb R^M$ and the linear map $A \in \mathbb{R}^{K \times M}$, with $M = 2$ and $K = 5$. The likelihood function is 
\begin{equation}
\label{eq:likelihood}
p(y | x) = \frac{1}{(\sqrt{2\pi} \sigma)^K} \exp\left(- \frac{\| y - Ax \|^2 }{2\sigma^2} \right).
\end{equation} 
If we assume the prior distribution of $x$ is a Gaussian distribution with mean $\bm{0}$ and covariance $S$, we can get an analytical expression for the posterior distribution of $x$ given the observation $y$,
\begin{equation}
    \label{eq:linear_posterior}
    p(x|y) = {\cal{C}} \exp \left(- \frac{1}{2} (y - \Sigma^{-1} b)^T \Sigma^{-1}  (y - \Sigma^{-1} b) \right), 
    \end{equation}
where
\begin{equation}
 {\cal{C}} = \frac{1}{\sqrt{(2\pi)^M | \Sigma | }}, \quad \Sigma = S^{-1} + \frac{1}{\sigma^2}A^TA, \quad b = \frac{1}{\sigma^2}A^Ty\,.\nonumber
\end{equation}
During the training stage we draw 
100 samples of $x$ from its prior $p(x)$, and $y$ is generated 
through the linear observation model~\eqref{eq:linear_obs}. We train 
a 3-layer model with the model objective~\eqref{model_obj_batch}, 
and show three examples of the posterior estimation in 
Figure~\ref{sim_no_variable_fig}. Therein we show 50\% and 90\% 
confidence contours. Black lines represent ground truth results  
(ellipses given the posterior is Gaussian), while the red 
contours correspond to the neural network estimations, based 
on Gaussian kernel density estimation (KDE) with 9,000 samples 
generated from the network. These results indicate that our deep learning 
model can produce reliable and statistically valid results.

\newcolumntype{A}{>{\centering}p{0.14\textwidth}}
\newcolumntype{C}{>{\centering\arraybackslash}p{0.12\textwidth}}
\begin{table*}
	\renewcommand{\arraystretch}{2.0}
	\caption{Data-driven and Bayesian results~\cite{o1o2catalog,abbott2020gwtc} for the median and 90\% confidence intervals of the masses of five binary black hole mergers. The network signal-to-noise ratio (SNRs) for each event are also provided for reference.}
	\label{real_event_values}
	\vskip -0.25in
	\begin{center}
		\begin{small}
			\begin{sc}
	    	\begin{tabular}{C||AA|AA|C}
					\hline
					Event Name & \multicolumn{2}{c|}{Our model} & \multicolumn{2}{c|}{LIGO~\cite{o1o2catalog,abbott2020gwtc}} & SNR \\
		\cmidrule(l){2-3} \cmidrule(l){4-5} 
            & $m_1 [\msun]$ & $m_2 [\msun]$ &$m_1 [\msun]$ & $m_2 [\msun]$ &\\
					\hline
					GW150914 & $38.85_{-4.15}^{+6.90}$ & $31.20_{-5.94}^{+4.39}$ & $35.60_{-3.10}^{+4.70}$ &  $30.60_{-4.40}^{+3.00}$ & 24.4 \\
					GW170104 & $28.90_{-3.80}^{+6.55}$ & $22.75_{-5.14}^{+3.73}$ & $30.80_{-5.60}^{+7.80}$ &  $20.00_{-4.60}^{+4.90}$ & 13.0 \\
					GW170814 & $33.92_{-5.27}^{+9.14}$ & $24.31_{-5.46}^{+4.13}$ & $30.60_{-5.30}^{+5.60}$ &  $25.20_{-4.00}^{+2.80}$ & 15.9 \\
					GW190521 & $46.10_{-6.61}^{+8.77}$ & $33.74_{-8.47}^{+6.68}$ & $42.10_{-4.90}^{+5.90}$ &  $32.70_{-6.20}^{+5.40}$ & 14.4 \\
					GW190630 & $34.00_{-4.43}^{+7.19}$ & $26.17_{-5.86}^{+4.54}$ & $35.00_{-5.70}^{+6.90}$ &  $23.60_{-5.10}^{+5.20}$ & 15.6 \\
					\hline
				\end{tabular}
			\end{sc}
		\end{small}
	\end{center}
	\vskip -0.1in
\end{table*}

\begin{table*}
	\renewcommand{\arraystretch}{2.0}
	\caption{Data-driven and Bayesian results~\cite{o1o2catalog,abbott2020gwtc} for the median and 90\% confidence intervals of the final spin of five binary black hole mergers. Results for the frequencies of the ringdown oscillations, \((\omega_I, \omega_R)\), are directly measure by our model from advanced LIGO's strain data, whereas the results quoted for LIGO are estimated using \(a_f\) values from~\cite{o1o2catalog,abbott2020gwtc} and  Eq.~\eqref{qnms}~\cite{Berti:2006}.}
	\label{real_event_values_omega1}
	\vskip -0.25in
	\begin{center}
		\begin{small}
			\begin{sc}
	    	\begin{tabular}{C||CCC|CCC}
					\hline
					Event Name & \multicolumn{3}{c|}{Our model} & \multicolumn{3}{c}{LIGO~\cite{o1o2catalog,abbott2020gwtc}} \\
		\cmidrule(l){2-4} \cmidrule(l){5-7}
            & $a_f$ & $\omega_R$ & $\omega_I$ &$a_f$ & $\omega_R$ & $\omega_I$\\
					\hline
					GW150914 & $0.71_{-0.07}^{+0.06}$ & $0.536_{-0.029}^{+0.028}$ & $0.0805_{-0.0026}^{+0.0023}$ & $0.69_{-0.04}^{+0.05}$ &  $0.528_{-0.023}^{+0.016}$ & $0.0811_{-0.0013}^{+0.0021}$ \\
					GW170104 & $0.69_{-0.07}^{+0.06}$ & $0.530_{-0.030}^{+0.028}$ & $0.0810_{-0.0025}^{+0.0023}$ & $0.66_{-0.11}^{+0.08}$ &  $0.515_{-0.033}^{+0.036}$ & $0.0821_{-0.0030}^{+0.0026}$ \\
					GW170814 & $0.68_{-0.09}^{+0.06}$ & $0.525_{-0.032}^{+0.028}$ & $0.0815_{-0.0024}^{+0.0024}$ & $0.72_{-0.05}^{+0.07}$ &  $0.541_{-0.022}^{+0.037}$ & $0.0800_{-0.0037}^{+0.0018}$ \\
					GW190521 & $0.73_{-0.06}^{+0.05}$ & $0.548_{-0.028}^{+0.029}$ & $0.0795_{-0.0029}^{+0.0024}$ & $0.72_{-0.07}^{+0.05}$ &  $0.552_{-0.030}^{+0.026}$ & $0.0800_{-0.0025}^{+0.0025}$ \\
					GW190630 & $0.71_{-0.07}^{+0.06}$ & $0.535_{-0.030}^{+0.028}$ & $0.0806_{-0.0026}^{+0.0024}$ & $0.70_{-0.07}^{+0.06}$ &  $0.532_{-0.028}^{+0.030}$ & $0.0808_{-0.0022}^{+0.0037}$ \\
					\hline
				\end{tabular}
			\end{sc}
		\end{small}
	\end{center}
	\vskip -0.1in
\end{table*}

\subsection{Results with Real Events}
\label{real_event_results_check}

In this section we use our deep learning models 
to estimate the medians 
and posterior distributions of the astrophysical 
parameters \((m_1, m_2)\) and \((a_f, \omega_R, \omega_I)\), 
respectively, for 
five binary black hole mergers: \texttt{GW150914}, \texttt{GW170104}, 
\texttt{GW170814}, \texttt{GW190521} and \texttt{GW190630}. 

As described in Section~\ref{related_work}, 
we consider 1s-long advanced LIGO noise input data 
batches, denoted as $y$, sampled at 4096Hz. We construct 
two posterior distribution estimations, 
$\hat{p}_{\omega, \theta}(x\vert y)$,  by minimizing 
the loss in Eq.~\eqref{model_obj_batch} for 
$(m_1, m_2)$ and for $(a_f, \omega_R, \omega_I)$. 
We use two different multivariate normal base 
distributions for $p(z)$ in the two different models. 
To estimate the masses of the binary components, 
the mean and covariance matrix ($\mu, \Sigma$) 
are: $\mu=(30, 30), \Sigma=\texttt{diag}(5, 5)$; whereas 
for the final spin and QNMs model we use: 
$\mu=(0.5, 0.55, 0.07), 
\Sigma=\texttt{diag}(0.05, 0.03, 0.002)$. 
``$\texttt{diag}(\cdot)$" refers to the diagonal matrix 
with ``$\cdot$" being the diagonal elements. The number 
of normalizing flow layers also varies for the two models. 
We use a 3-layer normalizing flow module for 
masses prediction, and an 8-layer module for the 
predictions of final spin and QNMs. 

\begin{figure*}
	\centering
	\subfigure[GW150914]{
		\includegraphics[width=0.40\linewidth]{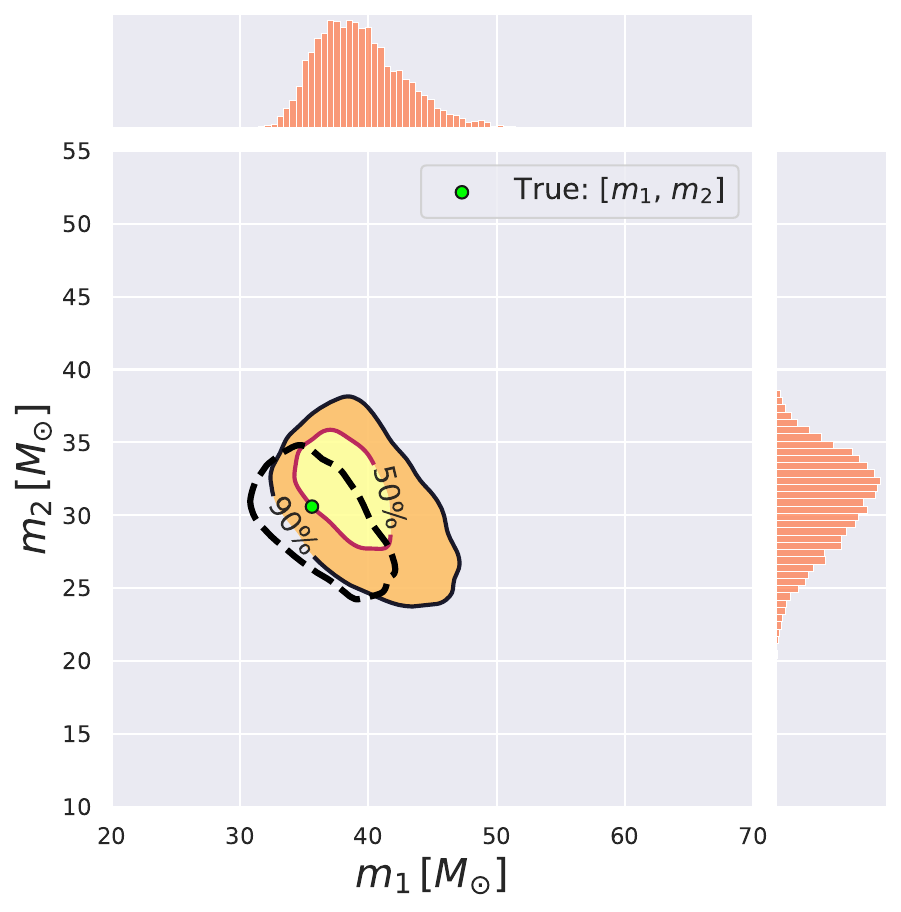}}
	\subfigure[GW170104]{
		\includegraphics[width=0.40\linewidth]{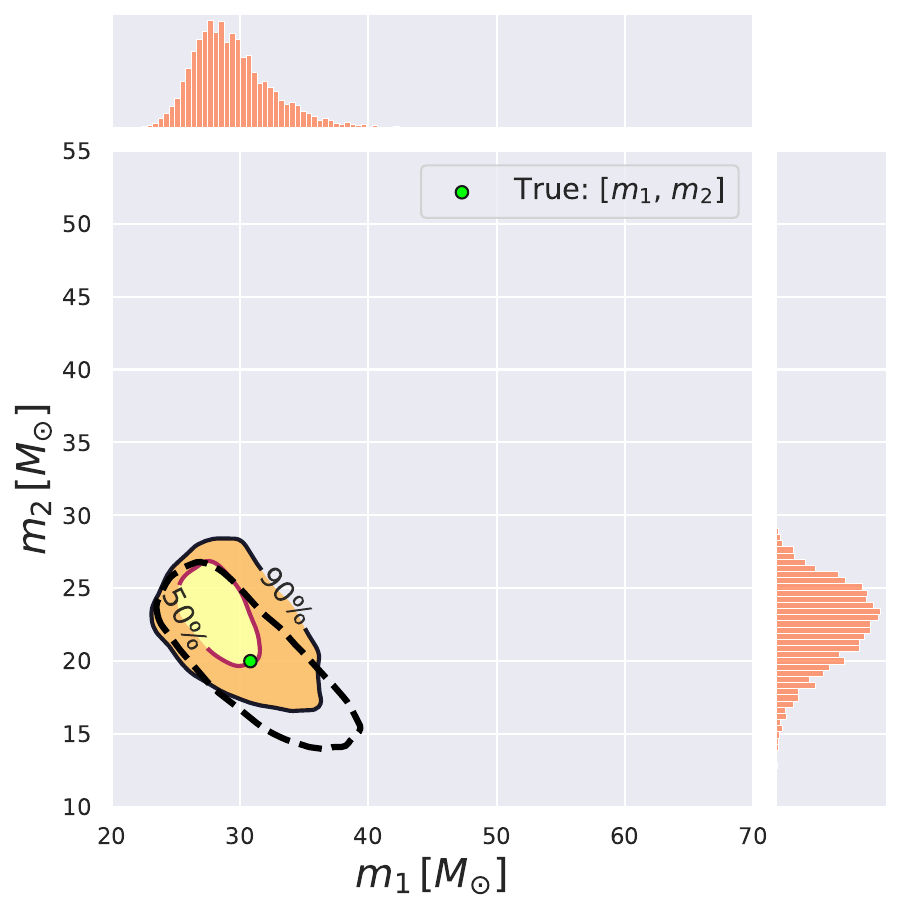}}
        \subfigure[GW170814]{
		\includegraphics[width=0.40\linewidth]{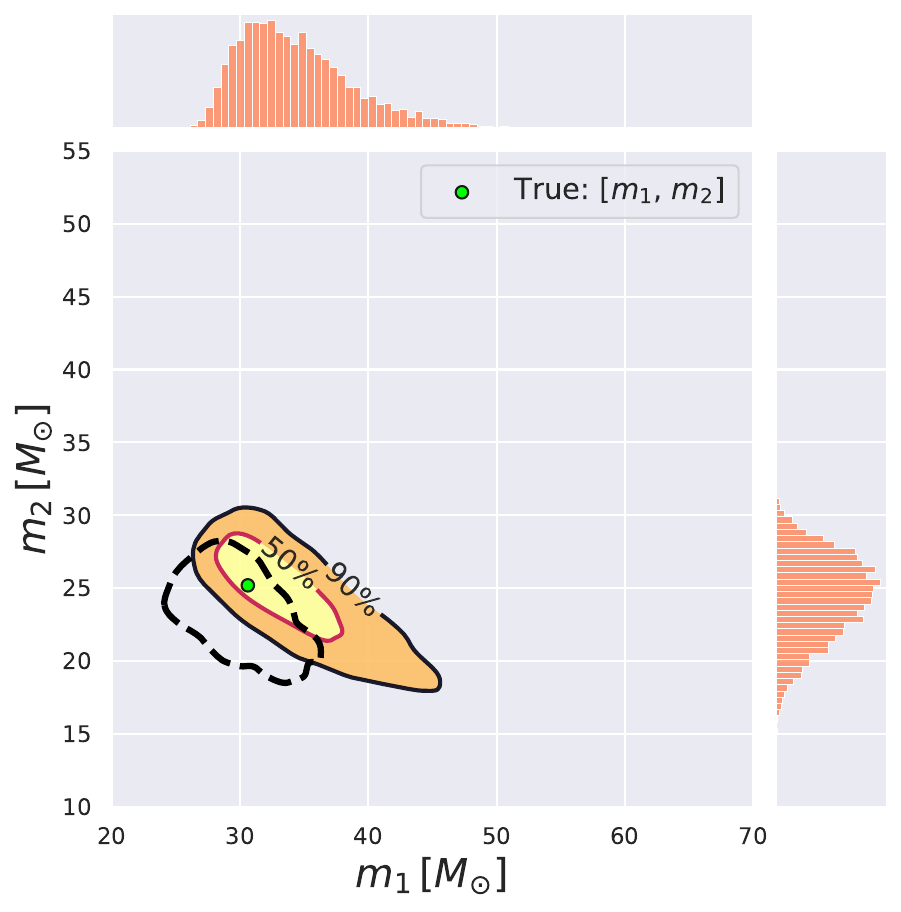}}
		\subfigure[GW190521]{
		\includegraphics[width=0.40\linewidth]{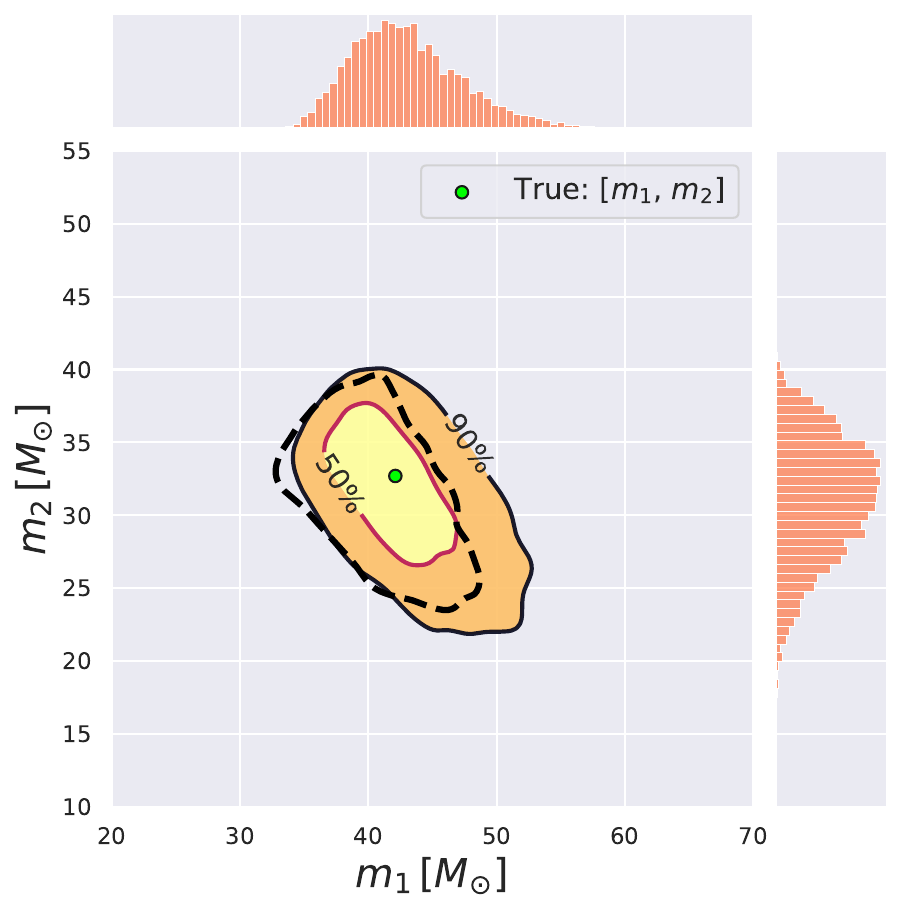}}
	    \subfigure[GW190630]{
		\includegraphics[width=0.40\linewidth]{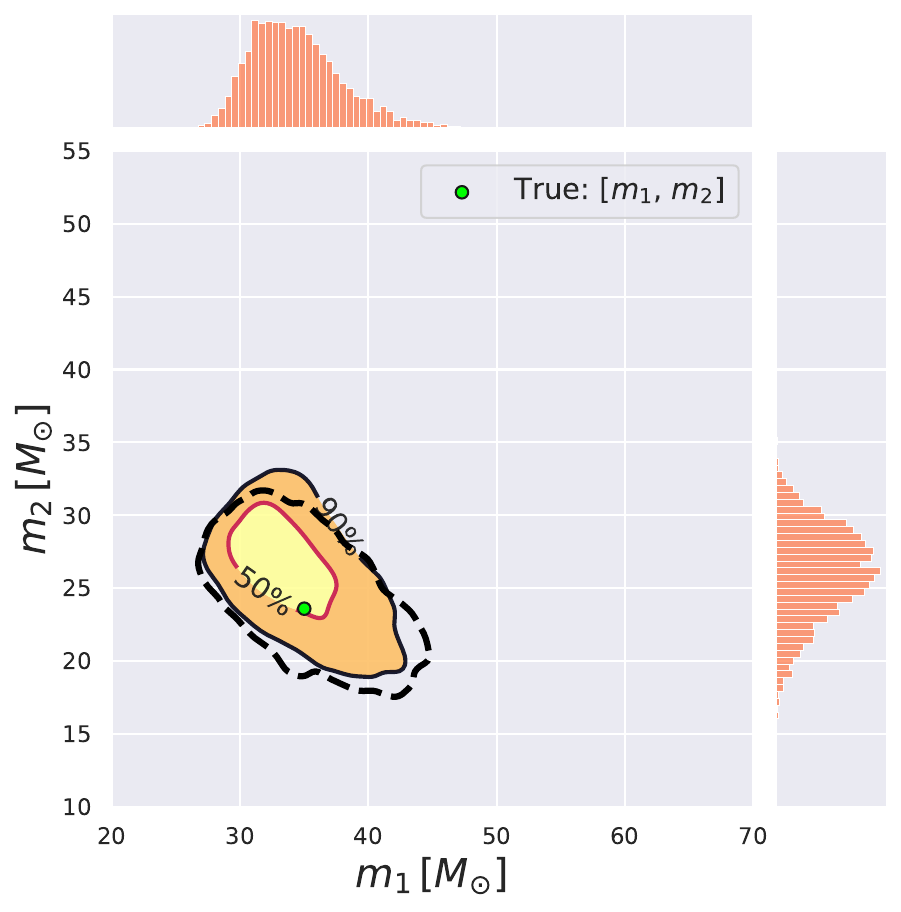}}
	\caption{Data-driven posterior distributions, including 50\% and 90\% confidence regions, for the masses of black hole mergers.}
	\label{contours_real_events}
\end{figure*}

\begin{figure*}
	\centering
	\subfigure[GW150914]{
		\includegraphics[width=0.40\linewidth]{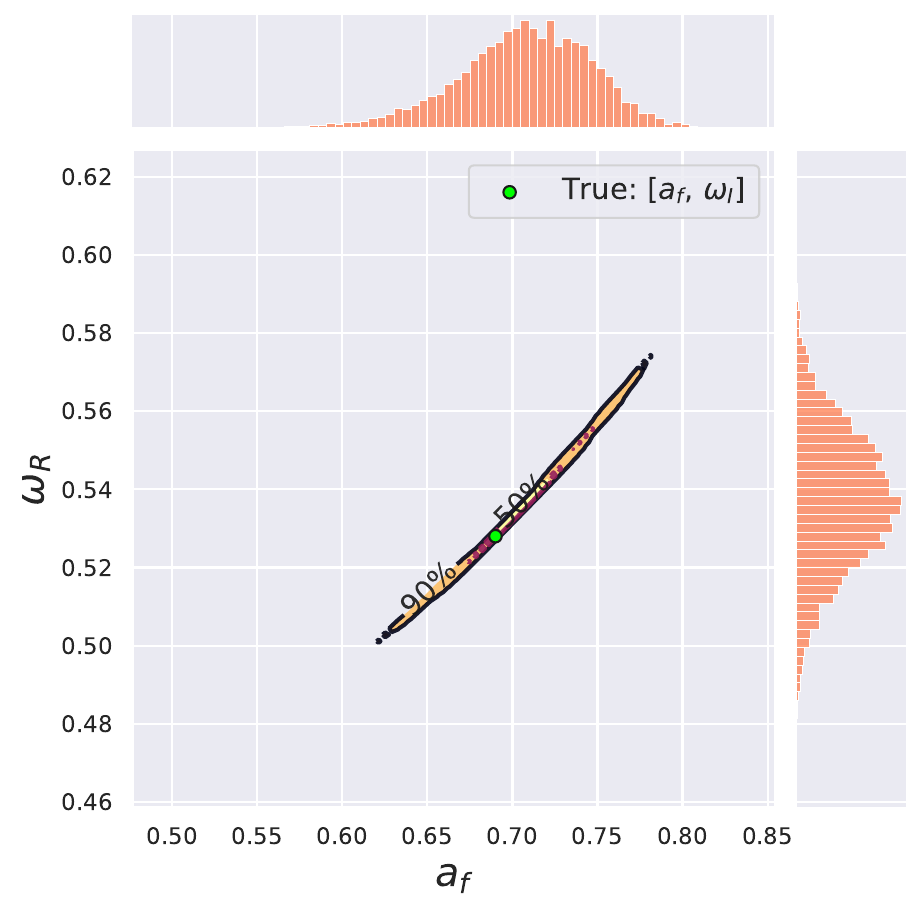}}
	\subfigure[GW170104]{
		\includegraphics[width=0.40\linewidth]{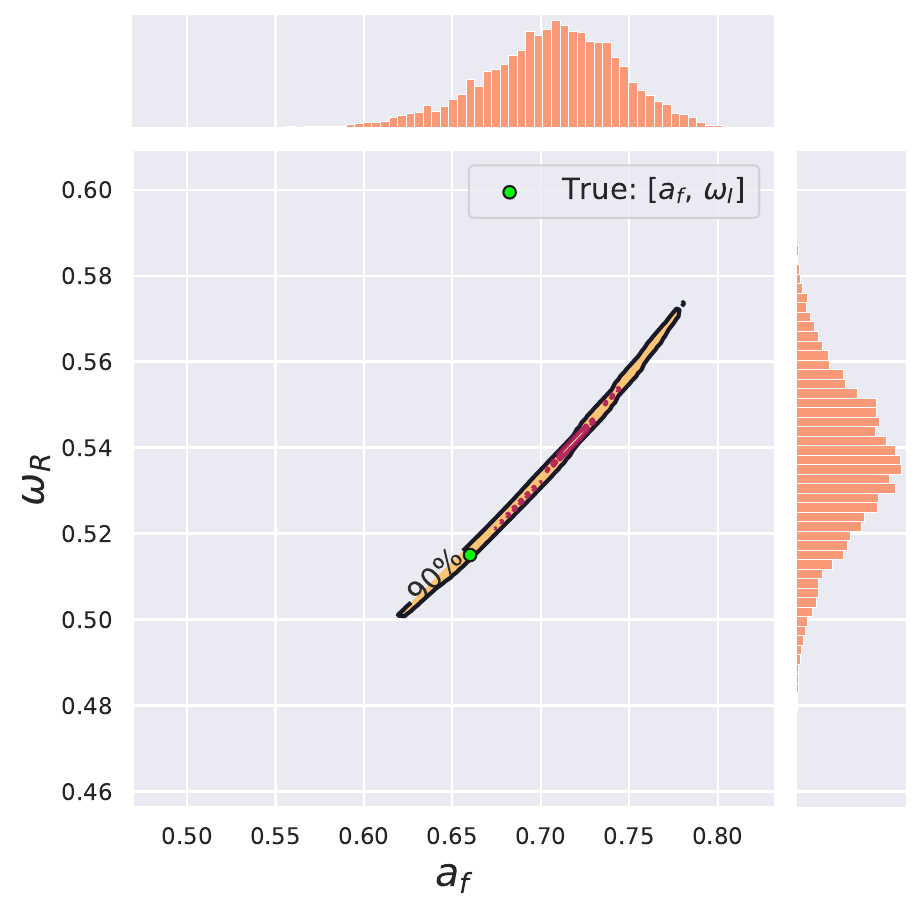}}
        \subfigure[GW170814]{
		\includegraphics[width=0.40\linewidth]{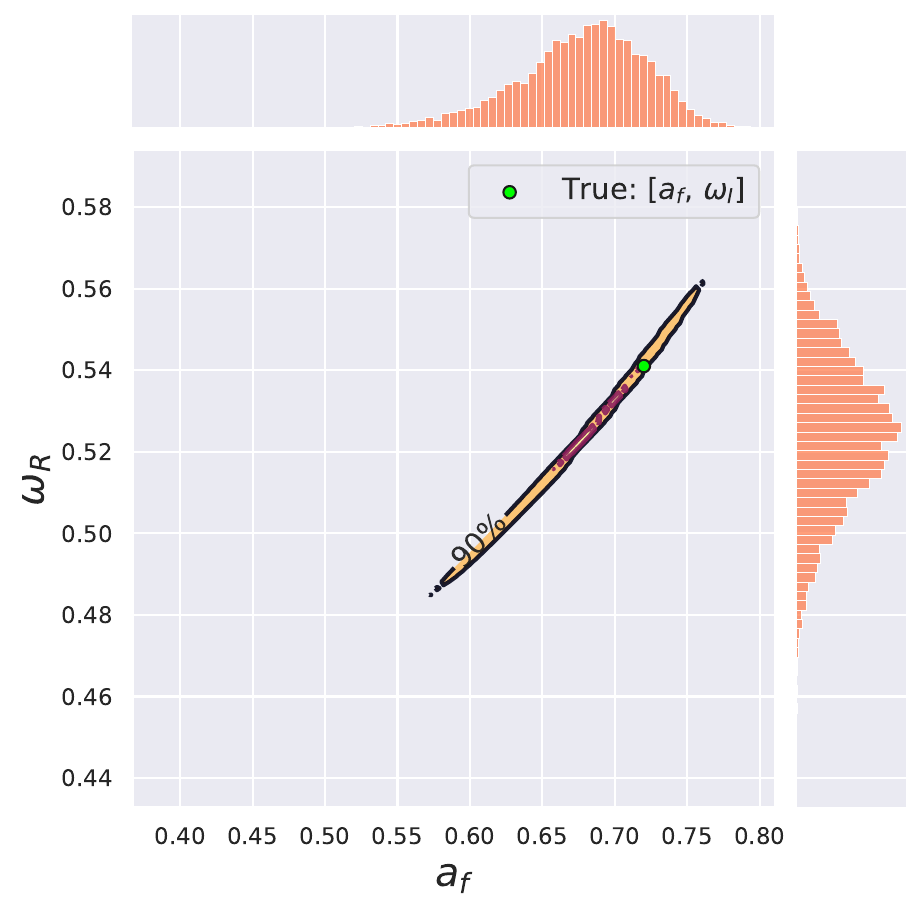}}
		\subfigure[GW190521]{
		\includegraphics[width=0.40\linewidth]{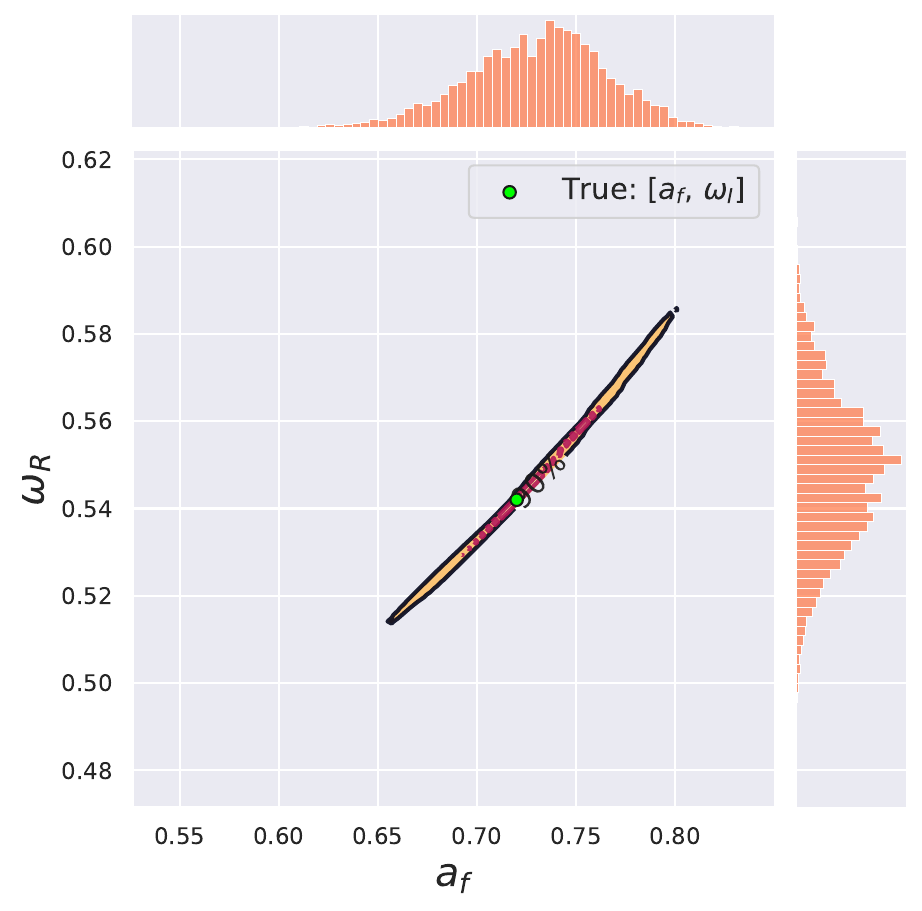}}
	    \subfigure[GW190630]{
		\includegraphics[width=0.40\linewidth]{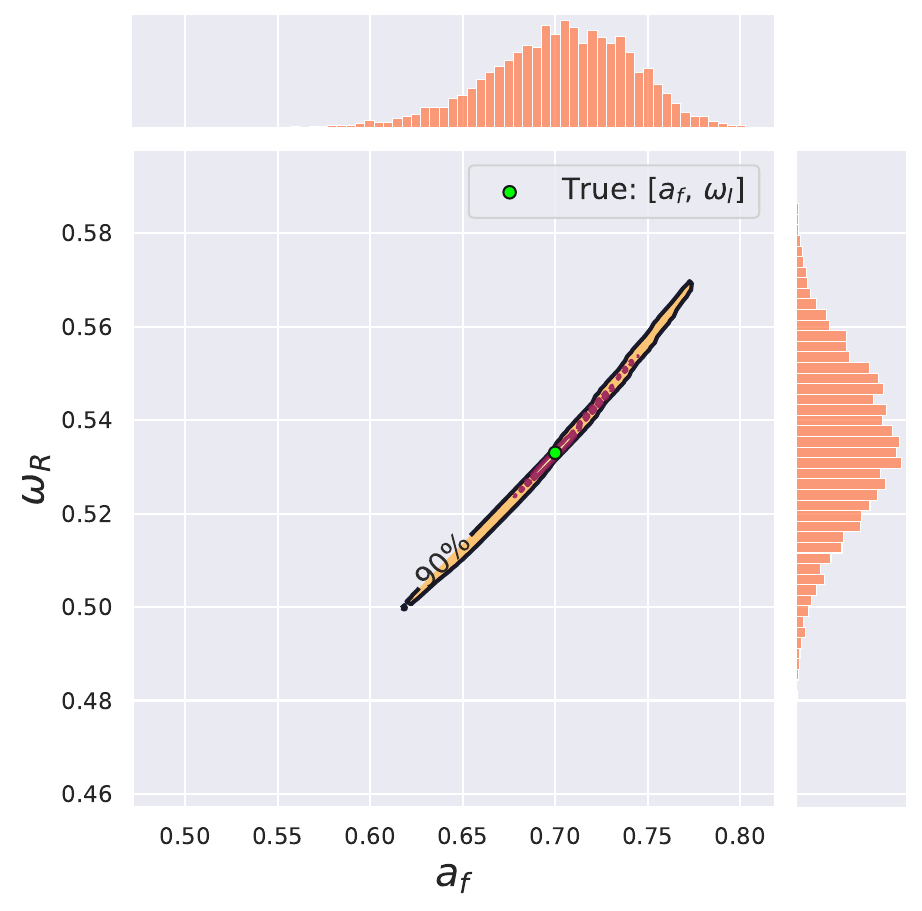}}
	\caption{Data-driven posterior distributions, including 50\% and 90\% confidence regions, for  $(a_f,\,\omega_R)$ of black hole mergers.}
	\label{contours_real_events_omega1}
\end{figure*}

\begin{figure*}
	\centering
	\subfigure[GW150914]{
		\includegraphics[width=0.40\linewidth]{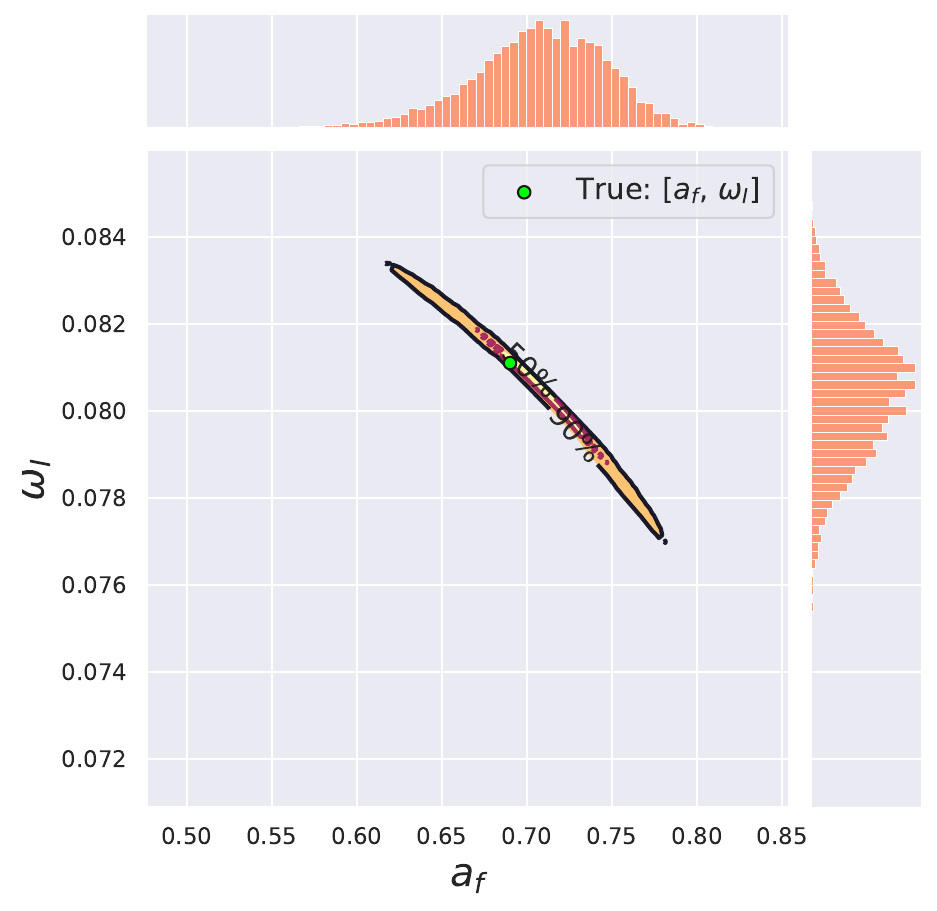}}
	\subfigure[GW170104]{
		\includegraphics[width=0.40\linewidth]{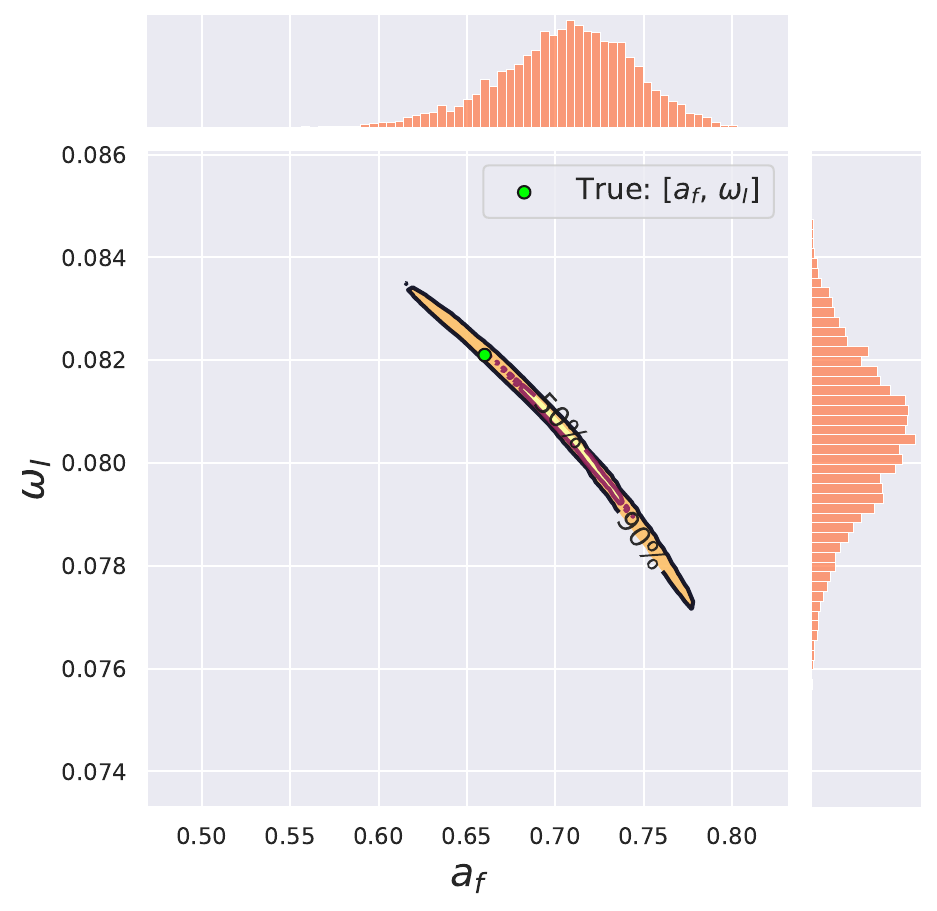}}
        \subfigure[GW170814]{
		\includegraphics[width=0.40\linewidth]{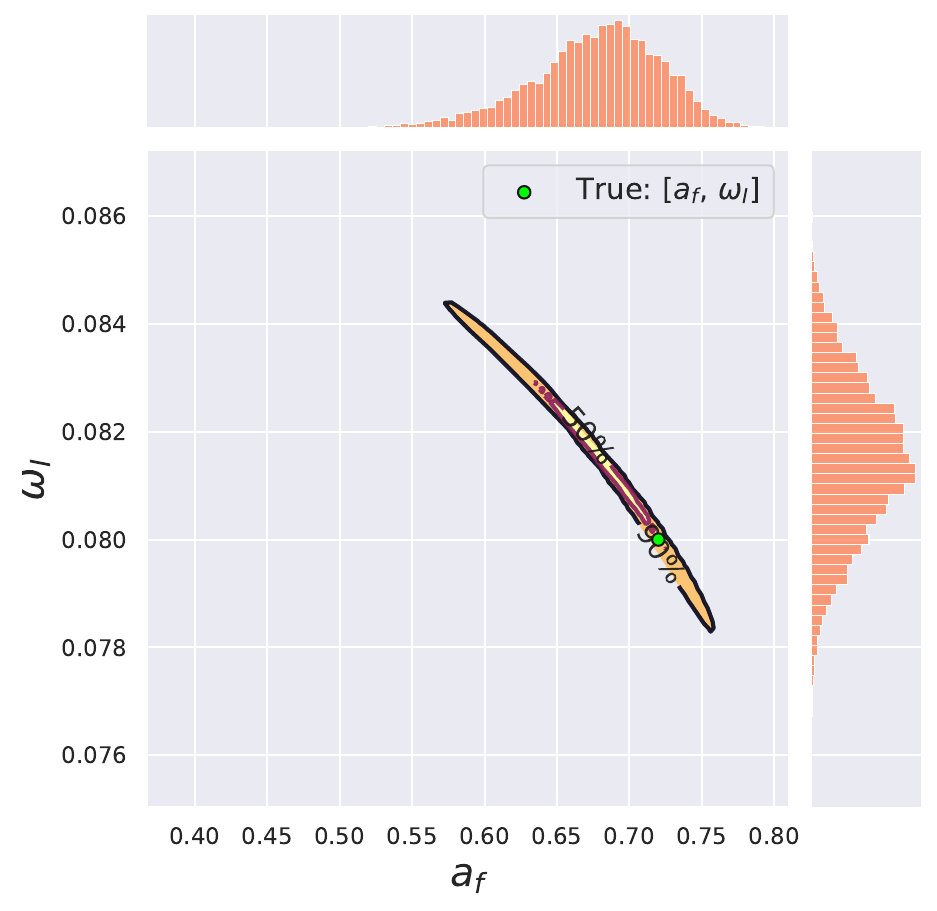}}
		\subfigure[GW190521]{
		\includegraphics[width=0.40\linewidth]{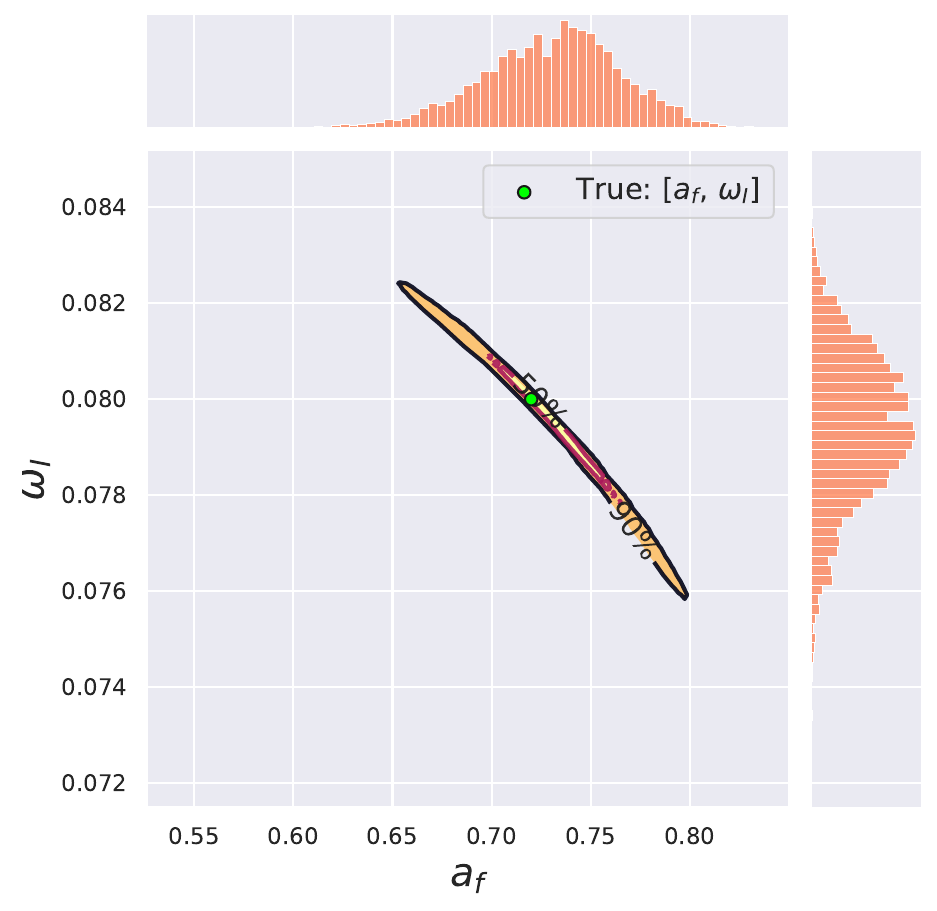}}
	    \subfigure[GW190630]{
		\includegraphics[width=0.40\linewidth]{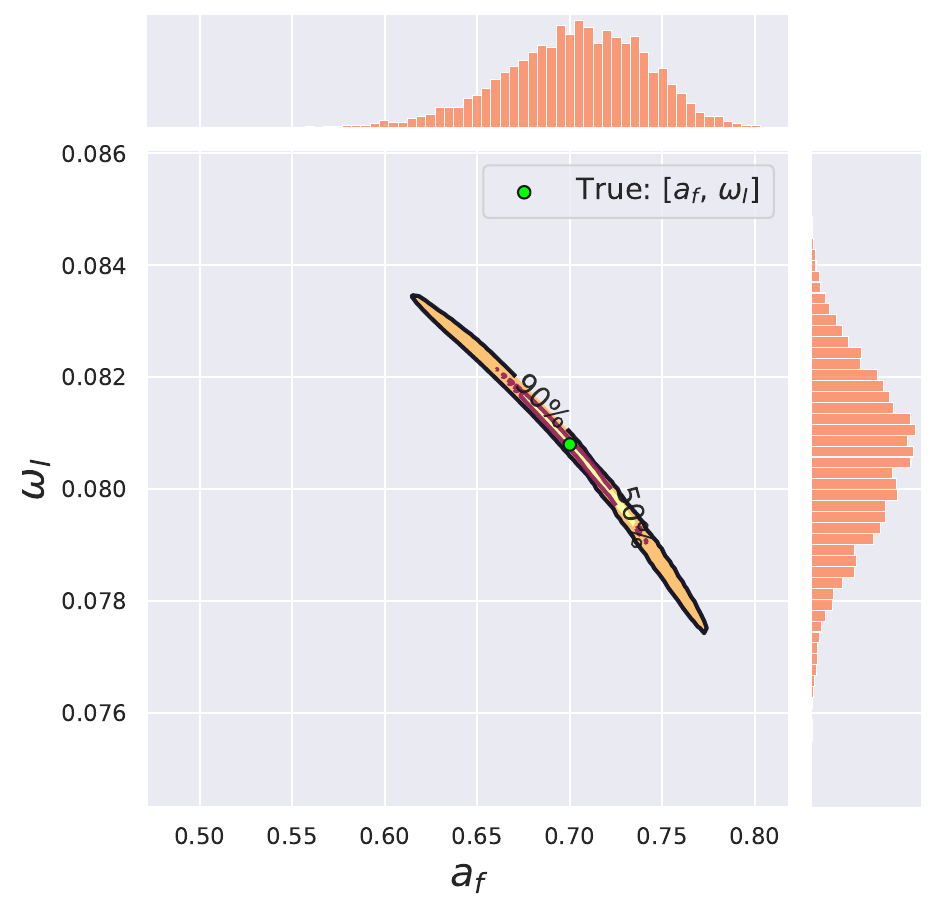}}
	\caption{Data-driven posterior distributions, including 50\% and 90\% confidence regions, for  $(a_f,\,\omega_I)$ of real black hole mergers.}
	\label{contours_real_events_omega2}
\end{figure*}

Our first set of results is presented in 
Figures~\ref{contours_real_events}, 
~\ref{contours_real_events_omega1}, 
and~\ref{contours_real_events_omega2}. These figures 
provide the median, and the 50\% and 90\% 
confidence intervals, which we computed using Gaussian 
KDE estimation with 9,000 samples drawn from the 
estimated posteriors. In Tables~\ref{real_event_values} and~\ref{real_event_values_omega1} we also present a 
summary of our data-driven median results and 
90\% confidence intervals, along with those obtained 
with traditional Bayesian algorithms in~\cite{o1o2catalog,abbott2020gwtc}.
Before we present the main highlights of 
these results, it is important to emphasize that our 
results are entirely data-driven. We have not attempted 
to use deep learning as a fast interpolator that 
learns the properties of traditional Bayesian posterior 
distributions. Rather, we have allowed deep learning to 
figure out the physical correlations among different 
parameters that describe the physics of black hole mergers. 
Furthermore, we have quantified the statistical consistency 
of our approach by validating it against a well known 
model. This is of paramount importance, since deep learning 
models may be constructed to reproduce the properties of 
traditional Bayesian distributions, but that fact does 
not provide enough evidence of their statistical validity 
or consistency. Finally, given the nature of the signal 
processing tools and computing approaches we 
use in this study, we do not expect our data-driven 
results to exactly reproduce the traditional Bayesian 
results reported in~\cite{o1o2catalog,abbott2020gwtc}. 

Our results may be summarized as 
follows. Figures~\ref{contours_real_events}, 
~\ref{contours_real_events_omega1}, 
and~\ref{contours_real_events_omega2} show that 
our data-driven posterior distributions encode expected physical correlations for the 
masses of the binary components, \((m_1,m_2)\), and the 
parameters of the remnant: \((a_f,\omega_R)\) and 
\((a_f,\omega_I)\). We also learn that these 
posterior distributions are determined by the 
properties of the 
noise and loudness of the signal that 
describes these events. \textcolor{black}{Figure~\ref{contours_real_events} presents a direct comparison between the posterior distributions predicted by our deep learning models and those produced with \texttt{PyCBC Inference}---marked with dashed lines. These results show that 
our deep learning models provide real-time, reliable information about the astrophysical properties of binary black hole mergers that were detected in three different observing runs, and which span a broad SNR range.}

\begin{figure}
	\centering
		\includegraphics[width=0.8\linewidth]{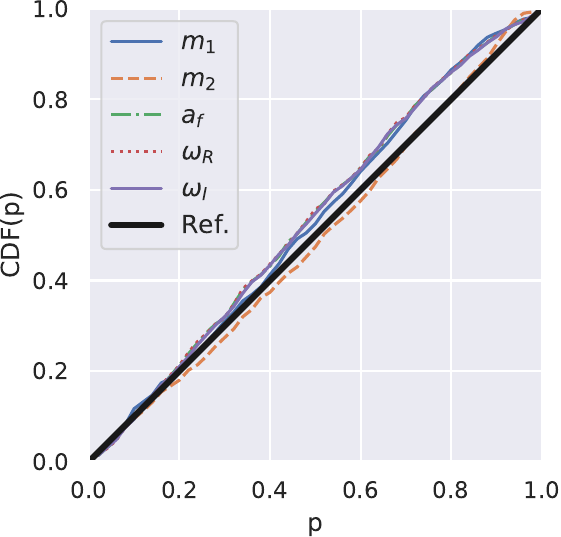}
	\caption{\textcolor{black}{P-P plot comparing the posterior distributions estimated by the neural network model for five astrophysical parameters \((m_1, m_2, a_f, \omega_R, \omega_I)\).}}
	\label{fig:pp_plot_LIGO}
\end{figure}

On the other hand, Tables~\ref{real_event_values} and~\ref{real_event_values_omega1} show that our 
median and 90\% confidence intervals are better, similar 
and in some cases slightly larger than those 
obtained with Bayesian algorithms. In these Tables, 
Bayesian LIGO results for \(a_f\) are directly taken from~\cite{o1o2catalog,abbott2020gwtc}, while 
\((\omega_R, \omega_I)\) results are computed using their Bayesian results for \(a_f\) and the tables available at~\cite{Berti:2005P}. These results indicate that deep learning methods can \textit{learn physical correlations in the data}, and 
provide reliable estimates of the parameters of 
gravitational wave sources. \textcolor{black}{
To demonstrate that our model represents true statistical properties of the posterior distribution, we tested the posterior estimation on simulated noisy gravitational waveforms. We calculate the empirical cumulative distribution function (CDF) of the number of times the true value for each parameter was found
within a given confidence interval $p$, as a function of $p$. We compare the empirical CDF with the true CDF of $p$ in the P-P plot in Figure~\ref{fig:pp_plot_LIGO}. To obtain the empirical CDF, for each test waveform (1000 waveforms in total) and one-dimensional
estimated posterior distribution generated from the network with 9,000 samples, we record the count of the confidence intervals $p$ ($p$=1\% , \dots, 100\%) where the true parameters fall. The empirical CDF is based on the frequency of such counts with the 1000 waveforms randomly drawn from the test dataset.  Since the empirical CDFs lie close to the diagonal, we conclude that the networks generate close approximation of the posteriors.} Furthermore, our data-driven 
results, including medians and posterior distributions, 
can be produced \textit{within 2 milliseconds per event} 
using a single \texttt{NVIDIA} V100 GPU. We expect that 
these tools will provide the means to assess in 
real-time whether the inferred astrophysical parameters 
of the binary components and the post-merger remnant 
adhere to general relativistic predictions. If not,
these results may prompt follow up analyses to 
investigate whether apparent discrepancies are due 
to poor data quality or other 
astrophysical effects~\cite{Gair:2012nm}.

 The reliable astrophysical information inferred in low-latency by deep learning algorithm warrants the extension of this framework to characterize other sources, including eccentric compact binary mergers, and sources 
 that require the inclusion of higher-order waveform modes. 
 Furthermore, the use of physics-inspired deep learning 
 architectures and optimization 
 schemes~\cite{Khan:2020foe} may enable an 
 accurate measurement of the spin of binary components. 
 These studies should be pursued in the future.

\section{Conclusion}
\label{conclusion}

We designed neural networks to estimate 
five parameters that describe the 
astrophysical properties of binary black holes 
before and after the merger event. The first two 
parameters constrain the masses of the 
binary components, while the others estimate 
the properties of the black hole remnant, namely 
\((m_1, m_2, a_f, \omega_R, \omega_I)\). These models 
combine a \texttt{WaveNet} 
architecture with normalizing flow and 
contrastive learning to provide statistically 
consistent estimates for both simulated distributions, 
and real gravitational wave sources. 

Our findings indicate that deep learning can 
abstract physical correlations in complex 
data, and then provide reliable predictions for 
the median and 90\% confidence intervals for 
binary black holes that span a broad SNR range. 
Furthermore, while these models 
were trained using only advanced LIGO noise from the 
first observing run, they were capable 
of generalizing to binary black holes that were 
reported during the first, second and third observing runs. 

These models will be extended in future work to 
provide informative estimates for the spin of the 
binary components, including higher-order waveform 
modes to better model the physics of highly spinning 
and asymmetric mass-ratio black hole systems.

\section{Acknowledgements}
\noindent Neural network models  are available 
at the \texttt{Data and Deep Learning Hub 
for Science}~\cite{blaiszik_foster_2019,dlhub}. 
EAH, HS and ZZ gratefully acknowledge National Science 
Foundation (NSF) awards OAC-1931561 and OAC-1934757.
EOS and PK gratefully acknowledge NSF grants PHY-1912081
and OAC-193128, and the Sherman Fairchild Foundation. PK
also acknowledges the support of the Department of Atomic
Energy, Government of India, under project no. RTI4001.
This work 
utilized the Hardware-Accelerated Learning (HAL)  cluster, 
supported by NSF Major Research Instrumentation program, 
grant OAC-1725729, as well as the University of Illinois 
at Urbana-Champaign.  Compute resources were provided by 
XSEDE using allocation TG-PHY160053. This  work  made  use 
of  the  Illinois  Campus  Cluster,  a computing resource 
that is operated by the Illinois Campus  Cluster  Program  
(ICCP)  in  conjunction  with  the National Center for 
Supercomputing Applications and  which  is  
supported  by  funds  from  the  University of Illinois 
at Urbana-Champaign. This research used resources of the 
Argonne Leadership Computing Facility, which is a DOE 
Office of Science User Facility supported under 
Contract DE-AC02-06CH11357.  
This research also made use of LIGO Data Grid clusters at
the California Institute of Technology. This research used
data, software and/or web tools obtained from the LIGO Open
Science Center (https://gw-openscience.org), a service of
LIGO Laboratory, the LIGO Scientific Collaboration and the Virgo
Collaboration. LIGO is funded by the U.S. National Science
Foundation. Virgo is funded by the French Centre
National de Recherche Scientifique (CNRS), the Italian
Istituto Nazionale della Fisica Nucleare (INFN) and the
Dutch Nikhef, with contributions by Polish and Hungarian institutes.

\bibliography{reference,references,ref_two,book_references}
\bibliographystyle{iopart-num}

\newpage

\end{document}